\documentclass[aps,prd,twocolumn,showpacs,amsmath,amssymb]{revtex4-1}
\usepackage{amsmath}
\usepackage{graphicx}
\usepackage{subfigure}
\usepackage{epstopdf}
\usepackage{color}
\usepackage{multirow}
\usepackage{setspace}
\usepackage{overpic}
\usepackage{amssymb}
\usepackage[driverfallback=dvipdfm, bookmarksnumbered, pdfstartview=FitH,colorlinks,urlcolor=blue, citecolor=blue,linkcolor=blue] {hyperref}
\usepackage{lineno}
\usepackage{bm}
\usepackage{rotating}
\usepackage[utf8]{inputenc}
\usepackage{threeparttable}

\let\oldequation\equation
\let\oldendequation\endequation

\renewenvironment{equation}
{\linenomathNonumbers\oldequation}
{\oldendequation\endlinenomath}

\begin{document}
	
	\title{\bf \boldmath
		Measurements of branching fractions of $D^0\to K^- 3\pi^+2\pi^-$, $D^0\to K^- 2\pi^+\pi^-2\pi^0$ and $D^+\to K^- 3\pi^+\pi^-\pi^0$
	}
	
	\author{
		\begin{small}
			\begin{center}
				M.~Ablikim$^{1}$, M.~N.~Achasov$^{4,c}$, P.~Adlarson$^{76}$, X.~C.~Ai$^{81}$, R.~Aliberti$^{35}$, A.~Amoroso$^{75A,75C}$, Q.~An$^{72,58,a}$, Y.~Bai$^{57}$, O.~Bakina$^{36}$, Y.~Ban$^{46,h}$
% ...
				\\
				\vspace{0.2cm}
				(BESIII Collaboration)\\
				\vspace{0.2cm} {\it
					$^{1}$ Institute of High Energy Physics, Beijing 100049, People's Republic of China\\
% ...
$^{4}$ Budker Institute of Nuclear Physics SB RAS (BINP), Novosibirsk 630090, Russia\\
% ...
$^{35}$ Johannes Gutenberg University of Mainz, Johann-Joachim-Becher-Weg 45, D-55099 Mainz, Germany\\
					$^{36}$ Joint Institute for Nuclear Research, 141980 Dubna, Moscow region, Russia\\
% ...
$^{46}$ Peking University, Beijing 100871, People's Republic of China\\
% ...
$^{57}$ Southeast University, Nanjing 211100, People's Republic of China\\
					$^{58}$ State Key Laboratory of Particle Detection and Electronics, Beijing 100049, Hefei 230026, People's Republic of China\\
% ...
$^{72}$ University of Science and Technology of China, Hefei 230026, People's Republic of China\\
% ...
$^{75}$ University of Turin and INFN, (A)University of Turin, I-10125, Turin, Italy; (B)University of Eastern Piedmont, I-15121, Alessandria, Italy; (C)INFN, I-10125, Turin, Italy\\
					$^{76}$ Uppsala University, Box 516, SE-75120 Uppsala, Sweden\\
% ...
$^{81}$ Zhengzhou University, Zhengzhou 450001, People's Republic of China\\
					\vspace{0.2cm}
					$^{a}$ Deceased\\
% ...
$^{c}$ Also at the Novosibirsk State University, Novosibirsk, 630090, Russia\\
% ...
$^{h}$ Also at State Key Laboratory of Nuclear Physics and Technology, Peking University, Beijing 100871, People's Republic of China\\
% ...
			}\end{center}
			\vspace{0.4cm}
		\end{small}
	}
	
	\begin{abstract}
		Utilizing $7.9\,\rm fb^{-1}$ of $e^+e^-$ collision data taken  with the BESIII detector at the center-of-mass energy of 3.773 GeV,
		we report the measurements of absolute branching fractions of the hadronic decays
		$D^0\to K^- 3\pi^+2\pi^-$, $D^0\to K^- 2\pi^+\pi^-2\pi^0$ and $D^+\to K^- 3\pi^+\pi^-\pi^0$.
		The $D^0\to K^- 3\pi^+2\pi^-$ decay is measured with improved precision, while the latter two decays are observed
		with statistical significance higher than $5\sigma$ for the first time.
		The absolute branching fractions of these decays are determined to be
		${\mathcal B}(D^0\to K^- 3\pi^+2\pi^-)=( 1.35\pm  0.23\pm  0.08 )\times 10^{-4}$,
		${\mathcal B}(D^0\to K^- 2\pi^+\pi^-2\pi^0)=( 19.0\pm  1.1\pm  1.5)\times 10^{-4}$,
		and
		${\mathcal B}(D^+\to K^- 3\pi^+\pi^-\pi^0)=( 6.57\pm  0.69\pm  0.33)\times 10^{-4}$,
		where the first uncertainties are statistical and the second systematic.
	\end{abstract}

	\maketitle
	
	\oddsidemargin  -0.2cm
	\evensidemargin -0.2cm
	
	\section{Introduction}
	
	Experimental measurements of hadronic $D$ decays are important for studies of CP violation, $D^0$-$\bar D^0$ mixing, and flavor SU(3) symmetry breaking effects
	in the charm sector~\cite{theory_1,theory_2,ref5,chenghy1,yufs}.
	Precise and comprehensive measurements of the absolute branching fractions (BFs) of
	hadronic decays of $D$ mesons containing three charged pions are valuable inputs for understanding important backgrounds in studies of $B \to D^{*-} \tau^+ \nu_\tau$ decays, where enticing hints of lepton flavor universality violation are observed~\cite{belle2-white-paper,lhcb-white-paper}.
	
	The measured BFs of the inclusive decays $D^{0(+)}\to \pi^+\pi^+\pi^-X$~\cite{bes3_D_3pix} indicate that there is some room for unmeasured $D^{0(+)}$ decays containing three charged pions, which are $(1.55\pm0.50)\%$ and $(0.51\pm0.57)\%$ for $D^0$ and $D^+$ decays,respectively. Certain Cabbibo-favored multi-body hadronic $D^{0(+)}$ decays containing
	one kaon and multiple pions, are promising, as proposed by Ref.~\cite{rosner1}.
	Much progress has been made recently in experimental studies of $D^{0(+)}$
	decays containing one kaon along with one to four pions~\cite{pdg}.
	However, the experimental knowledge of the hadronic $D^{0(+)}$ decays
	containing one kaon and five pions is very poor,
	mainly due to limited data sample and small BFs~\cite{pdg}.
	Previously, only the FOCUS Collaboration reported ${\cal B}
	(D^0\to K^- 3\pi^+2\pi^-) = ( 2.2\pm  0.6)\times 10^{-4}$~\cite{kppppp}
	based on a relative branching ratio.
	The large $e^+e^-$ collision data sample taken at the $\psi(3770)$ peak with
	the BESIII detector allows for a measurement of the absolute BFs
	of these six-body hadronic decays.
	
	In this paper, we report measurements of the absolute BFs of the hadronic decays $D^0\to K^- 3\pi^+2\pi^-$, $D^0\to K^- 2\pi^+\pi^-2\pi^0$ and $D^+\to K^- 3\pi^+\pi^-\pi^0$,
	by analyzing the $e^+e^-$ collision data corresponding to an integrated luminosity of 7.9~fb$^{-1}$~\cite{lum_bes31,lum_bes32,lum_bes33} collected with the BESIII detector at the
	center-of-mass energy $E_{\rm cm}=3.773$~GeV.
	Throughout this paper, charge conjugation is always implied,
	and $D \, (\bar{D})$ denotes $D^+, \, D^0 \, (\bar{D}^0, \, D^-)$ mesons.
	
	\section{BESIII detector and Monte Carlo simulation}
	
	The BESIII detector~\cite{Ablikim:2009aa} records symmetric $e^+e^-$ collisions
	provided by the BEPCII storage ring~\cite{Yu:IPAC2016-TUYA01}
	in the center-of-mass energy range from 1.84 to 4.95~GeV,
	with a peak luminosity of $1.1 \times 10^{33}\;\text{cm}^{-2}\text{s}^{-1}$
	achieved at $E_{\rm cm} = 3.773\;\text{GeV}$. The cylindrical core of the BESIII detector covers 93\% of the full solid angle and consists of a helium-based multilayer drift chamber~(MDC), a time-of-flight system~(TOF), and a CsI(Tl) electromagnetic calorimeter~(EMC),
	which are all enclosed in a superconducting solenoidal magnet
	providing a 1.0~T magnetic field.
	The solenoid is supported by an
	octagonal flux-return yoke with resistive plate counter muon
	identification modules interleaved with steel.
	The charged-particle momentum resolution at $1~{\rm GeV}/c$ is
	$0.5\%$, and the
	${\rm d}E/{\rm d}x$
	resolution is $6\%$ for electrons
	from Bhabha scattering. The EMC measures photon energies with a
	resolution of $2.5\%$ ($5\%$) at $1$~GeV in the barrel (end-cap)
	region. The time resolution in the plastic scintillator TOF barrel region is 68~ps, while
	that in the end-cap region was 110~ps.
	The end-cap TOF
	system was upgraded in 2015 using multigap resistive plate chamber
	technology, providing a time resolution of
	60~ps,
	which benefits 63\% of the data used in this analysis~\cite{etof}.
	
	Monte Carlo (MC) simulated data samples produced with a {\sc
		geant4}-based~\cite{geant4} software package, which
	includes the geometric description of the BESIII detector and the
	detector response, are used to determine detection efficiencies
	and to estimate backgrounds. The simulation models the beam
	energy spread and initial state radiation (ISR) in the $e^+e^-$
	annihilations with the generator {\sc
		kkmc}~\cite{ref:kkmc}.
	The inclusive MC sample includes the production of $D\bar{D}$
	pairs (with quantum coherence for the $D^0, \bar{D}^0$ channels),
	the non-$D\bar{D}$ decays of the $\psi(3770)$, the ISR
	production of the $J/\psi$ and $\psi(3686)$ states, and the
	continuum processes incorporated in {\sc kkmc}~\cite{ref:kkmc}.
	All particle decays are modeled with {\sc
		evtgen}~\cite{ref:evtgen} using BFs
	either taken from the
	Particle Data Group~\cite{pdg}, when available,
	or otherwise estimated with {\sc lundcharm}~\cite{ref:lundcharm}.
	Final state radiation
	from charged final state particles is incorporated using the {\sc
		photos} package~\cite{photos2}.
	
	\section{Measurement Method}
	
	In $e^+e^-$ collisions at $E_{\rm cm}=3.773$~GeV,
	the $D^0\bar D^0$ or $D^+D^-$ pairs are produced without any additional hadrons.
	This property offers an ideal platform to measure the absolute BFs of the hadronic $D$ decays by using the double-tag~(DT) method~\cite{doubletag}.
	The single-tag~(ST) $\bar D^0$ mesons are reconstructed from three hadronic decays $\bar D^0 \to K^+\pi^-$, $K^+\pi^-\pi^0$, $K^+\pi^-\pi^-\pi^+$, and
	the ST $D^-$ mesons are reconstructed from six hadronic decays $D^- \to K^{+}\pi^{-}\pi^{-}$,
	$K^0_{S}\pi^{-}$, $K^{+}\pi^{-}\pi^{-}\pi^{0}$, $K^0_{S}\pi^{-}\pi^{0}$, $K^0_{S}\pi^{+}\pi^{-}\pi^{-}$, and $K^{+}K^{-}\pi^{-}$. Then DT candidates are formed by selecting signal decays in the recoiling side against the $\bar{D}$ mesons.
	The BF of the signal decay can be determined as
	\begin{equation}
		\label{eq:br}
		{\mathcal B}_{{\rm sig}} = N_{\rm DT}/(N^{\rm tot}_{\rm ST} \, \epsilon_{{\rm sig}}),
	\end{equation}
	where $N^{\rm tot}_{\rm ST}$ is the yield of ST $\bar D$ mesons summed over all tag modes,
	$N_{\rm DT}$ is the yield of DT events, and
	$\epsilon_{{\rm sig}}$ is the efficiency of detecting the signal $D$ decay, averaged over all tag modes, given by
	\begin{equation}
		\label{eq:eff}
		\epsilon_{{\rm sig}} = \sum_i (N^i_{{\rm ST}} \, \epsilon^i_{{\rm DT}}/\epsilon^i_{{\rm ST}})/N^{\rm tot}_{\rm ST},
	\end{equation}
	where $\epsilon^i_{{\rm ST}}$ and $\epsilon^i_{{\rm DT}}$ are the efficiencies of detecting ST and DT candidates in the tag mode $i$, respectively.
	
	\section{Event selection}
	
	All charged tracks, except those from $K^0_{S}$ decays, are required to satisfy $ V_{xy} <1$~cm, $|V_{z}|<10$~cm,
	where $V_{xy}$ and $V_z$ are the distances of closest approach to the interaction point along the beam direction
	and in the plane perpendicular to the beam direction, respectively.
	We also require $\vert\!\cos\theta\vert<0.93$, where $\theta$ is the polar angle with respect to the symmetry axis of the MDC.
	We perform particle identification (PID) on charged tracks with combined $dE/dx$ and TOF information, using the calculated confidence levels for the pion and kaon hypotheses, $CL_{\pi}$ and $CL_{K}$.
	Tracks with $CL_{K}>CL_{\pi}$ and $CL_{\pi}>CL_{K}$ are assigned as kaon and pion candidates, respectively.
	
	The $K^0_S$ candidates are reconstructed via $K^0_S\to \pi^+\pi^-$ decays.
	Two oppositely charged tracks are required to satisfy $|V_{z}|<20$~cm and $\vert\!\cos\theta\vert<0.93$, and they are assumed to be pions with no PID.
	The accepted $\pi^+\pi^-$ pairs are then constrained to originate from a common vertex and their invariant mass is required to be within $(0.487,0.511)~{\rm GeV}/c^2$,
	corresponding to approximately three times the fitted resolution around the known $K^0$ mass~\cite{pdg}.
	The decay length of each $K^0_S$ candidate must be at least twice the vertex resolution from the interaction point.  The four-vector from the vertex fit is used for later kinematics.
	
	The $\pi^0$ candidates are reconstructed via $\pi^0\to \gamma\gamma$ decays. Each photon candidate is selected from EMC shower start time within 700~ns of the event start time and a minimal deposited energy of more than
	25~MeV in the barrel region ($|\!\cos \theta\vert< 0.80$) or 50~MeV in the end-cap region ($0.86 <|\!\cos \theta\vert< 0.92$).
	The energy deposited in the neighboring TOF counters is included to improve the reconstruction efficiency and energy resolution. The minimum opening angle between the shower and the nearest charged track must be larger than $10^{\circ}$.
	The $\gamma\gamma$ combinations with invariant masses in the range of $(0.115,\,0.150)$\,GeV$/c^{2}$ are retained as $\pi^0$ candidates.
	To improve the resolution, a kinematic fit is performed on the selected photon pair, constraining the $\gamma\gamma$ invariant mass to the known $\pi^{0}$ mass~\cite{pdg} and the constrained four-vector is used for later kinematics.
	
	For $\bar D^0\to K^+\pi^-$ candidates, backgrounds from cosmic rays and Bhabha events are removed with the following requirements. First, the two charged tracks must have a TOF time difference of less than 5~ns and they must not be consistent with being a muon pair or an electron-positron pair. Second, there must be at least one EMC shower with an energy larger than 50 MeV or at least one additional good ({\it i.e.}, passing the previous selections) charged track detected in the MDC~\cite{deltakpi}.
	
	\section{Yields of single-tag $\bar D$ mesons}
	
	We use the same method to obtain the ST yields and ST efficiencies as in Ref.~\cite{STyields}.
	To distinguish the tagged $\bar D$ mesons from combinatorial backgrounds,
	we define two kinematic variables, {\it e.g.}, energy difference
	\begin{equation}
		\Delta E_{\rm tag} \equiv E_{\rm tag} - E_{\rm b},
		\label{eq:deltaE}
	\end{equation}
	and beam-constrained mass
	\begin{equation}
		M_{\rm BC}^{\rm tag} \equiv \sqrt{E^{2}_{\rm b}-|\vec{p}_{\rm tag}|^{2}},
		\label{eq:mBC}
	\end{equation}
	where $E_{\rm b}$ is the beam energy, and $\vec{p}_{\rm tag}$ and $E_{\rm tag}$
	are the momentum and energy of the tagged $\bar D$ candidate in the rest frame of $e^+e^-$ system, respectively.
	The $\Delta E^{i}$ requirements for different tag modes are listed in Table~\ref{ST:realdata}.
	For each tag mode, if there are multiple candidates in an event,
	we only retain the one giving the minimum value of $|\Delta E_{\rm tag}|$~\cite{STyields}.
	
	\begin{table*}
		\centering
		\caption {The $\Delta E^{i}$ requirements, the ST $\bar D$ yields in data, $N^i_{\rm ST}$, and the ST efficiencies, $\epsilon_{\rm ST}^{i}$, for each tag mode, where the uncertainties are statistical only.}
		
		\begin{tabular}{lccc}
			\hline
			\hline
			{Tag mode \hskip 1.8cm} & {\hskip 0.5cm $\Delta E^{i}$~(GeV) \hskip 0.5cm}  &  {\hskip 0.3cm$ N^{i}_{\rm ST}~(\times 10^3)$ \hskip 0.3cm} &  $\epsilon^i_{\rm ST}~(\%)$       \\\hline
			$\bar D^{0}\to K^{+}\pi^{-}$                           & $(-0.027,0.027)$&$1482.8\pm1.3$&     $66.89\pm0.01$ \\
			$\bar D^{0}\to K^{+}\pi^{-}\pi^{0}$                   & $(-0.062,0.049)$&$3118.2\pm2.1$&      $37.68\pm0.01$ \\
			$\bar D^{0}\to K^{+}\pi^{-}\pi^{-}\pi^{+}$            & $(-0.026,0.024)$ &$1997.9\pm1.6$&     $41.88\pm0.01$ \\
			\hline
			$D^{-}\to K^{+} \pi^{-}\pi^{-}$                       & $(-0.025,0.024)$&$2215.3\pm1.6$&     $52.44\pm0.01$ \\
			$D^{-}\to K_S^{0} \pi^{-}$                            & $(-0.025,0.026)$&$256.0\pm0.5$&      $51.89\pm0.02$ \\
			$D^{-}\to K^{+}\pi^{-}\pi^{-}\pi^{0}$                 & $(-0.057,0.046)$&$735.4\pm1.2$&      $27.19\pm0.01$ \\
			$D^{-}\to K_S^{0} \pi^{-}\pi^{0}$                     & $(-0.062,0.049)$&$590.4\pm1.0$&      $27.57\pm0.01$ \\
			$D^{-}\to K_S^{0} \pi^{+}\pi^{-}\pi^{-}$              & $(-0.028,0.027)$& $307.7\pm0.7$&     $29.68\pm0.01$ \\
			$D^{-}\to K^{+}K^{-}\pi^{-}$                          & $(-0.024,0.023)$&$191.8\pm0.5$&      $42.05\pm0.02$ \\
			\hline
			\hline
		\end{tabular}
	\label{ST:realdata}
\end{table*}

\begin{figure}[htp]
	\centering
	\includegraphics[width=1.0\linewidth]{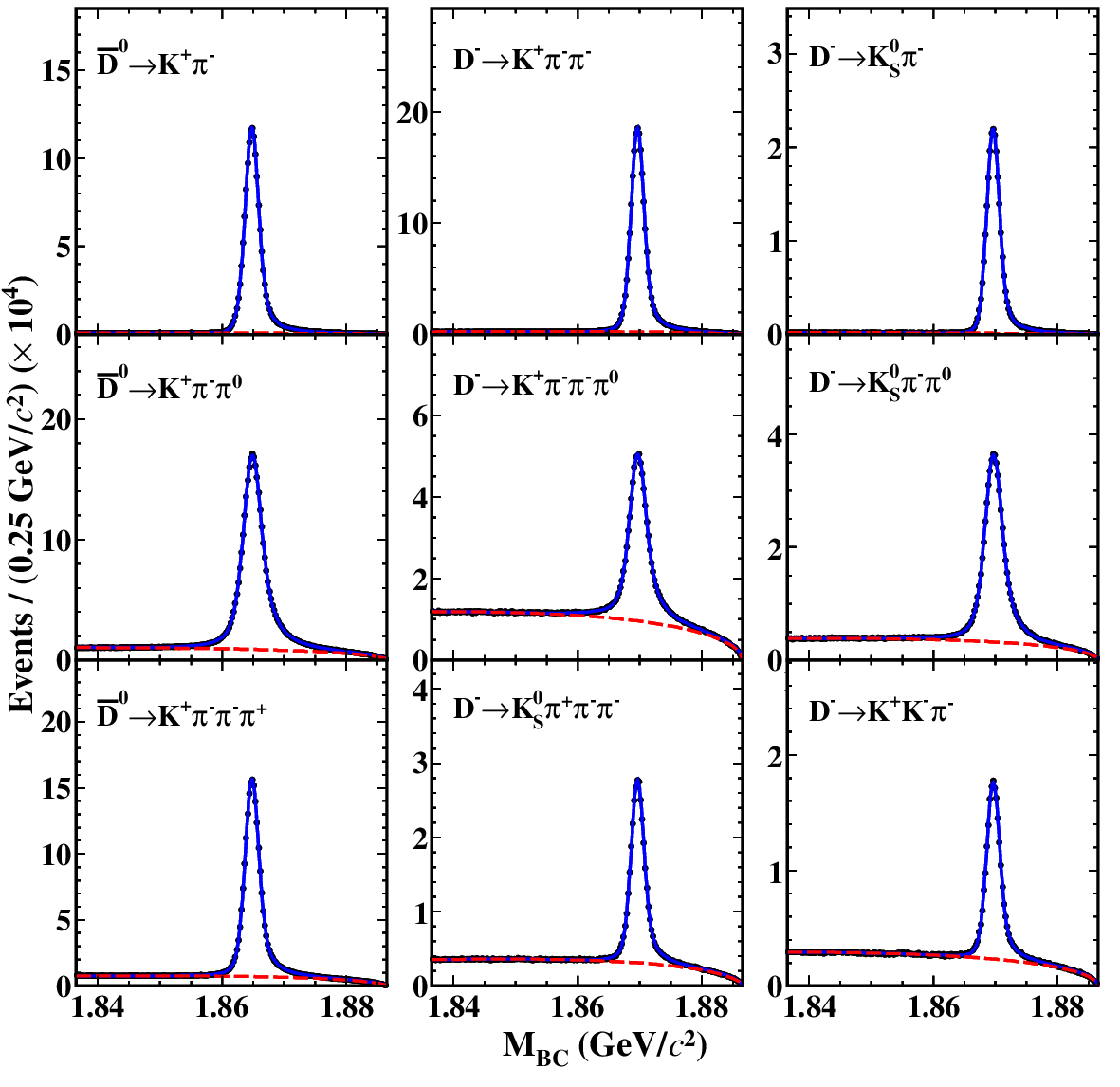}
	\caption{\small
		Fits to the $M_{\rm BC}$ distributions of the $\bar D^0$ (left column)
		and $D^-$ (middle and right columns) ST candidates.
		The points with error bars are data,
		the solid blue curves are the fit results,
		and the dashed red curves are the fitted background components.
	}
	\label{fig:datafit_MassBC}
\end{figure}

To extract the yield of ST $\bar D$ mesons for each tag mode, a binned maximum-likelihood fit is performed on the
$M_{\rm BC}^{\rm tag}$ distribution of the accepted candidates.
In the fit, the $\bar D$ signal is described with an MC-simulated shape convolved with
a double-Gaussian function to take into account the difference in resolution between data and MC simulation.
The combinatorial background is described by an ARGUS function~\cite{ARGUS}.
The results of the fits to the $M_{\rm BC}$ distributions of different tag modes are shown in Fig.~\ref{fig:datafit_MassBC}.
The ST $\bar D$ yields in data and ST efficiencies for the different tag modes are listed in Table~\ref{ST:realdata}.
Summing over all tag modes gives the total yields of ST $\bar D^0$ and $D^-$ mesons to be $(6599.0 \pm 2.9_{\rm stat.})\times 10^3$ and $(4296.6\pm2.4_{\rm stat.})\times 10^3$, respectively.

\section{Yields of double-tag events}

Candidates for signal $D$ decays are selected using the remaining tracks and showers not used for reconstructing the tag side $\bar{D}$ (ST).
The charged $D$ signal decay must have charge opposite to that of the tag side;
while for the neutral $D$ signal decays, the charge of the kaon on the signal side must be opposite to that on the tag side.
For all three signal decays, to suppress backgrounds with $\pi^+\pi^-$
produced from $K^0_S\to \pi^+\pi^-$ decays, the $\pi^+\pi^-$ invariant masses
are required to be outside the range of $(0.468,0.528)$~GeV/$c^2$.
For $D^0\to K^- 2\pi^+\pi^-2\pi^0$, to suppress backgrounds from
$D^0 \rightarrow K^-2\pi^+\pi^-K^0_S(\to \pi^0\pi^0)$,
the $\pi^0\pi^0$ invariant masses are required to be outside the range
of $(0.448,0.548)$~GeV/$c^2$.
We note that various narrow hadronic resonances (e.g., $\eta, \eta^\prime, \omega, \phi$) potentially present in the final state are not removed.

The signal $D$ mesons are identified using the energy difference $\Delta E_{\rm sig}$
and the beam-constrained mass $M_{\rm BC}^{\rm sig}$, which are calculated with Eq.~\ref{eq:deltaE} and Eq.~\ref{eq:mBC}, respectively,
by replacing ``sig'' with ``tag''.
For each signal $D$ decay, if there are multiple candidates in an event, only the one with the smallest $|\Delta E_{\rm sig}|$ is kept  for further analysis.
The signal $D$ decays are required to satisfy mode-dependent $\Delta E_{\rm sig}$ requirements,
as shown in the second column of Table~\ref{tab:DT}.

\begin{figure}[htp]
	\centering
	\includegraphics[width=1.0\linewidth]{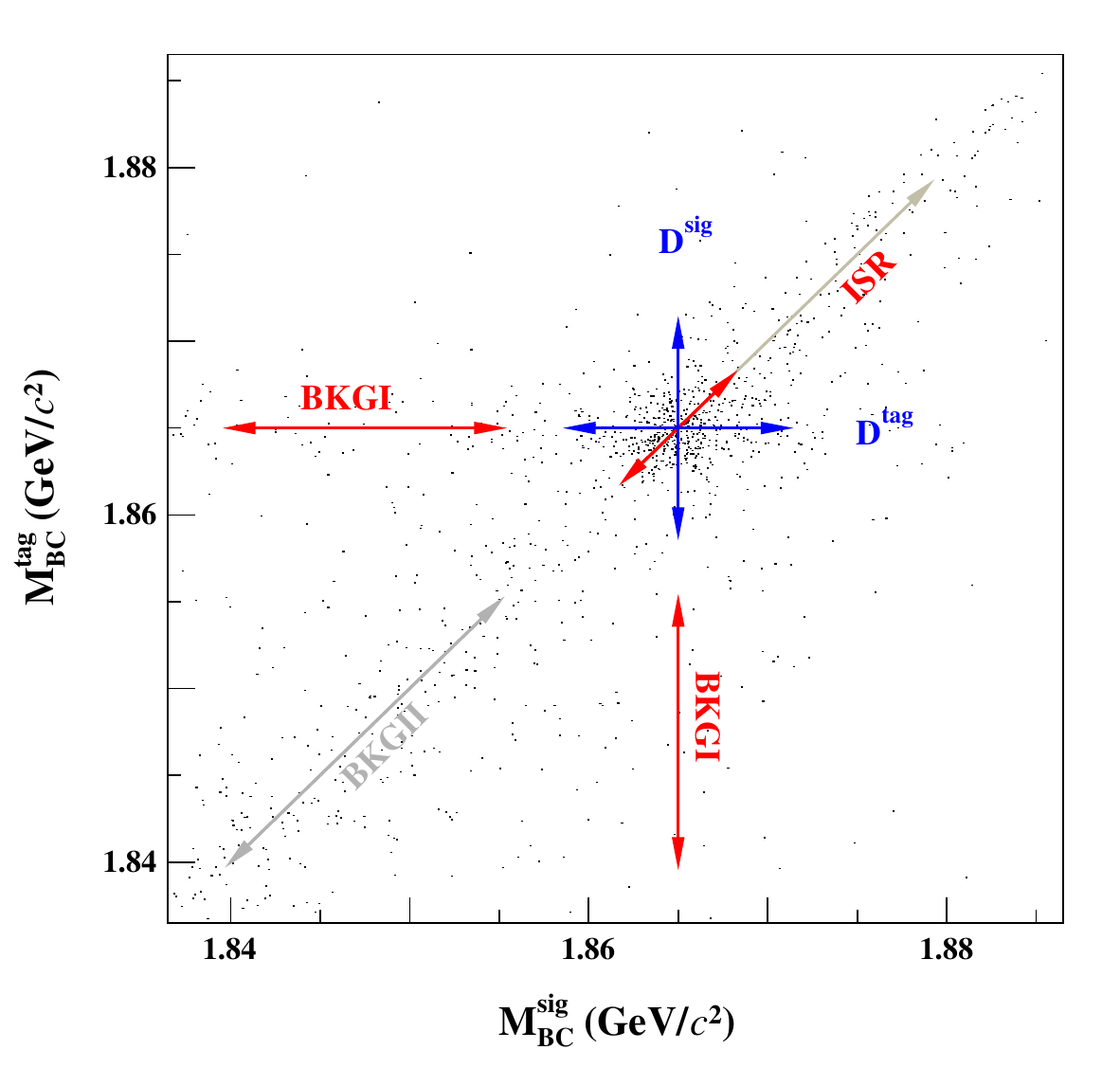}
	\caption{
		The distribution of $M_{\rm BC}^{\rm tag}$
		versus $M_{\rm BC}^{\rm sig}$ of the candidates for $D^0\to K^- 2\pi^+\pi^-2\pi^0$ in data, which have been summed over all $\bar D^0$ tag modes.
		Here, $D^{\rm sig}$ and $D^{\rm tag}$ denote the bands from correctly-reconstructed signals, with $M_{\rm BC}^{\rm sig} = M_{D}$, and correctly-reconstructed tags, with $M_{\rm BC}^{\rm tag} = M_{D}$,  respectively.
		Backgrounds are discussed in the main text.
	}
	\label{fig:mBC2D}
\end{figure}

Figure~\ref{fig:mBC2D} shows the $M_{\rm BC}^{\rm tag}$
versus $M_{\rm BC}^{\rm sig}$ distribution of the accepted DT candidates in data.
The signal events concentrate around $M_{\rm BC}^{\rm tag} = M_{\rm BC}^{\rm sig} = M_{D}$,
where $M_{D}$ is the known $D$  mass~\cite{pdg}.
The events distributed along the lines near $M_{\rm BC}^{\rm tag} = M_{D}$ or $M_{\rm BC}^{\rm sig} = M_{D}$, defined as BKGI, are mainly from a correctly reconstructed $D\,(\bar D)$ combined with an incorrectly reconstructed $\bar D\,(D)$.
The events smeared along the diagonal line, defined as BKGII, referred to as ISR, are mainly from the $e^+e^- \to q\bar q$ processes and incorrectly reconstructed $D\bar{D}$.
The events dispersed across the whole plane, defined as BKGIII, are mainly from incorrectly reconstructed $D$ and $\bar D$.

For each signal $D$ decay, the yield of DT events, $N^{\rm fit}_{\rm DT}$, is obtained from a two-dimensional (2D) unbinned maximum-likelihood
fit~\cite{cleo-2Dfit} to the $M_{\rm BC}^{\rm tag}$ versus $M_{\rm BC}^{\rm sig}$ distribution of the accepted candidates. In the fit, the probability
density functions (PDFs) of signal, BKGI, BKGII, and BKGIII are constructed as
\begin{itemize}
	\item
	Signal: $a(x,y)$,
	\item
	BKGI: $b(x) \, c_y(y;E_{\rm b},\xi_{y}) + b(y) \, c_x(x;E_{\rm b},\xi_{x})$,
	\item
	BKGII: $c_z(z;\sqrt{2}E_{\rm b},\xi_{z}) \c, g(k)$,
	\item
	BKGIII: $c_x(x;E_{\rm b},\xi_{x}) \, c_y(y;E_{\rm b},\xi_{y})$,
\end{itemize}
respectively.
Here, $x=M_{\rm BC}^{\rm sig}$, $y=M_{\rm BC}^{\rm tag}$, $z=(x+y)/\sqrt{2}$, and $k=(x-y)/\sqrt{2}$.
The PDFs of signal $a(x,y)$, $b(x)$, and $b(y)$ are taken from the corresponding MC-simulated shapes and $c_f(f;E_{\rm end},\xi_f)$ is the ARGUS function
defined as $A_f \, f\, (1 - \frac {f^2}{{E_{\rm b}}^2})^{0.5} \,\,  e^{\xi_{f}(1- (f/E_{\rm b})^2)}$,
where $f$ denotes $x$, $y$, or $z$; $E_{\rm b}$ is fixed at 1.8865 GeV,
$A_f$ is  a normalization factor;
and $\xi_f$ is a fit parameter. The signal shape $a(x, y)$ is convolved with a 2D Gaussian function. The PDF $g(k)$ is a Gaussian function with mean of zero and standard deviation parameterized by $\sigma_k=\sigma_0 \, (\sqrt{2}E_{\rm b}/c^2-k)^p$,
where $\sigma_0$ and $p$ are fit parameters.
For each signal decay, the statistical significance is greater than $5\sigma$,
as calculated from the maximum likelihoods with and without the signal
component in the fit and accounting for the change in
the number of degrees of freedom.

\begin{figure}[htbp]
	\centering
	\includegraphics[width=1.0\linewidth]{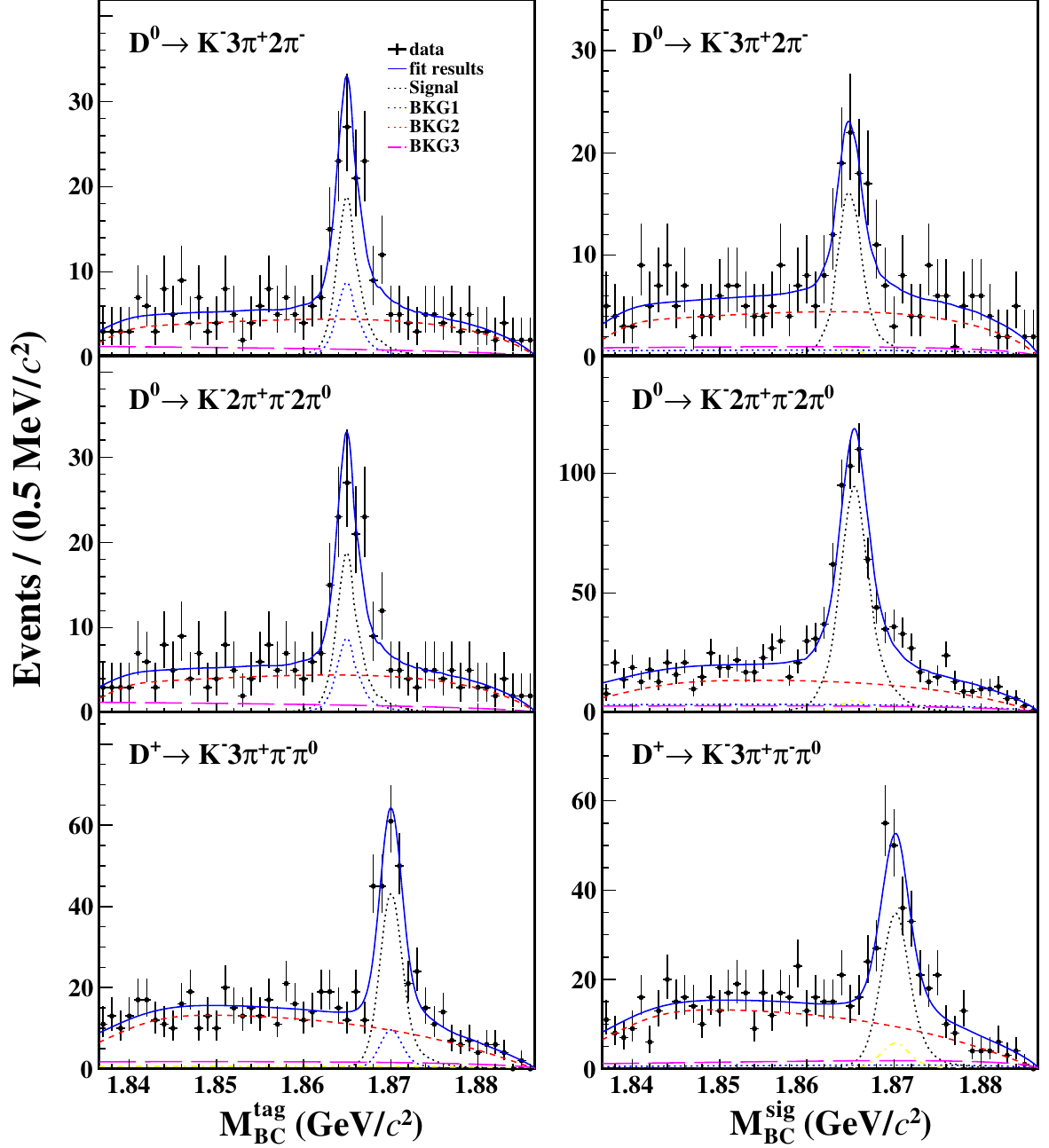}
	\caption{\small
		Projections of the 2D fits to the distributions of (left) $M^{\rm tag}_{\rm BC}$ and (right) $M^{\rm sig}_{\rm BC}$.
		The points with error bars are data.
		The solid blue, dotted black, dot-dashed blue, dot-long-dashed red, long-dashed magenta and dashed green curves denote the overall fit results,
		signal, BKGI, BKGII, BKGIII and peaking background components (see text), respectively.
	}
	\label{fig:2Dfit}
\end{figure}

Figure~\ref{fig:2Dfit} shows the $M^{\rm tag}_{\rm BC}$ and
$M^{\rm sig}_{\rm BC}$ projections of the 2D fits to the data.
From these fits, we obtain the DT yields of each signal $D$ decay;
these results are listed in Table~\ref{tab:DT}.

\begin{figure*}[!htp]
	\centering
	\includegraphics[width=0.9\linewidth]{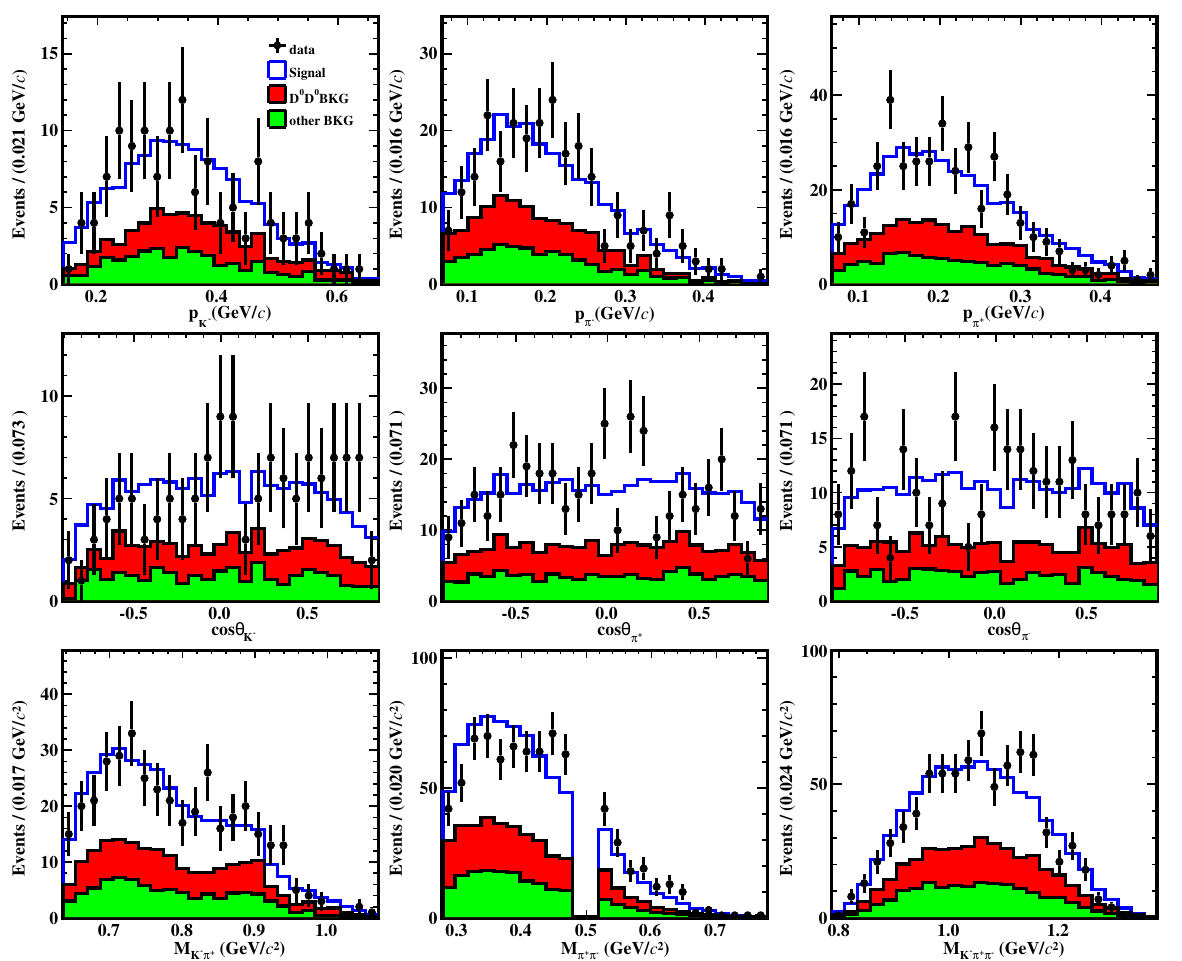}
	\caption{
		Comparisons of the distributions of momenta and cosines of polar angles of daughter particles,
		the invariant masses of two-body or three-body particle combinations of the accepted candidates for $D^0\to K^- 3\pi^+2\pi^-$
		between data (dots with error bars) and the total MC simulation (blue histogram). This total MC histogram is the sum of the signal MC events (white histogram) plus the MC-simulated backgrounds from the inclusive MC samples (red and green histograms).
		Events here satisfy the additional requirements of
		$|M_{\rm BC}^{\rm tag}-1.865|<0.005$~GeV/$c^2$ and $|M_{\rm BC}^{\rm sig}-1.865|<0.005$~GeV/$c^2$
		and there are three entries for $\pi^+$ and two entries for $\pi^-$ per event.
		\label{fig:compare1}
	}
\end{figure*}

\begin{figure*}[!htp]
	\centering
	\includegraphics[width=0.9\linewidth]{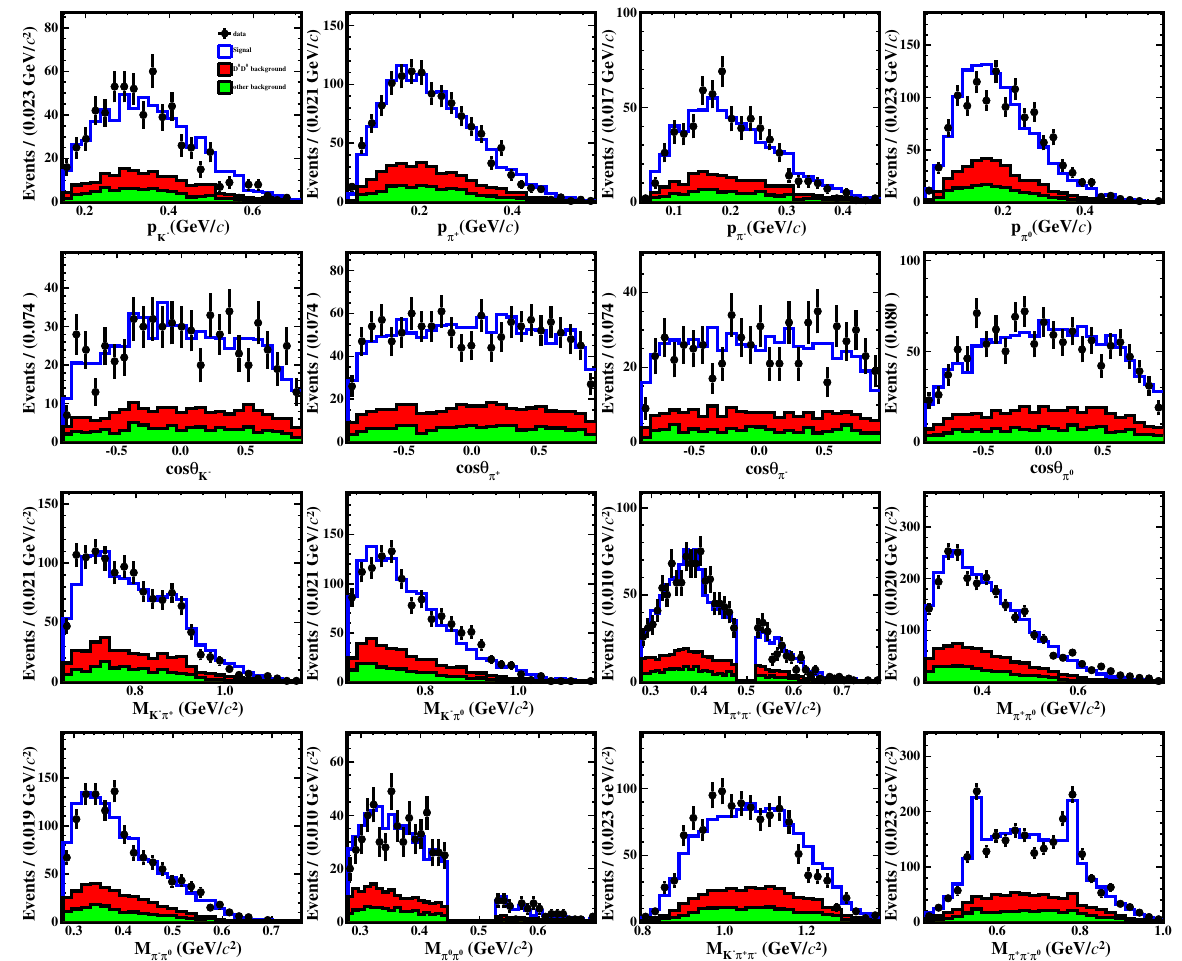}
	\caption{
		Comparisons of the distributions of momenta and cosines of polar angles of daughter particles,
		and the invariant masses of two-body or three-body particle combinations of the accepted candidates for $D^0\to K^- 2\pi^+\pi^-2\pi^0$
		between data (dots with error bars) and the total MC simulation (blue histogram). This total MC histogram is the sum of the signal MC events (white histogram) plus the MC-simulated backgrounds from the inclusive MC samples (red and green histograms).
		Events here satisfy the additional requirements of
		$|M_{\rm BC}^{\rm tag}-1.865|<0.005$~GeV/$c^2$ and $|M_{\rm BC}^{\rm sig}-1.865|<0.005$~GeV/$c^2$
		and there are two entries for $\pi^+$ and two entries for $\pi^0$ per event.
		\label{fig:compare2}
	}
\end{figure*}

\begin{figure*}[!htp]
	\centering
	\includegraphics[width=0.9\linewidth]{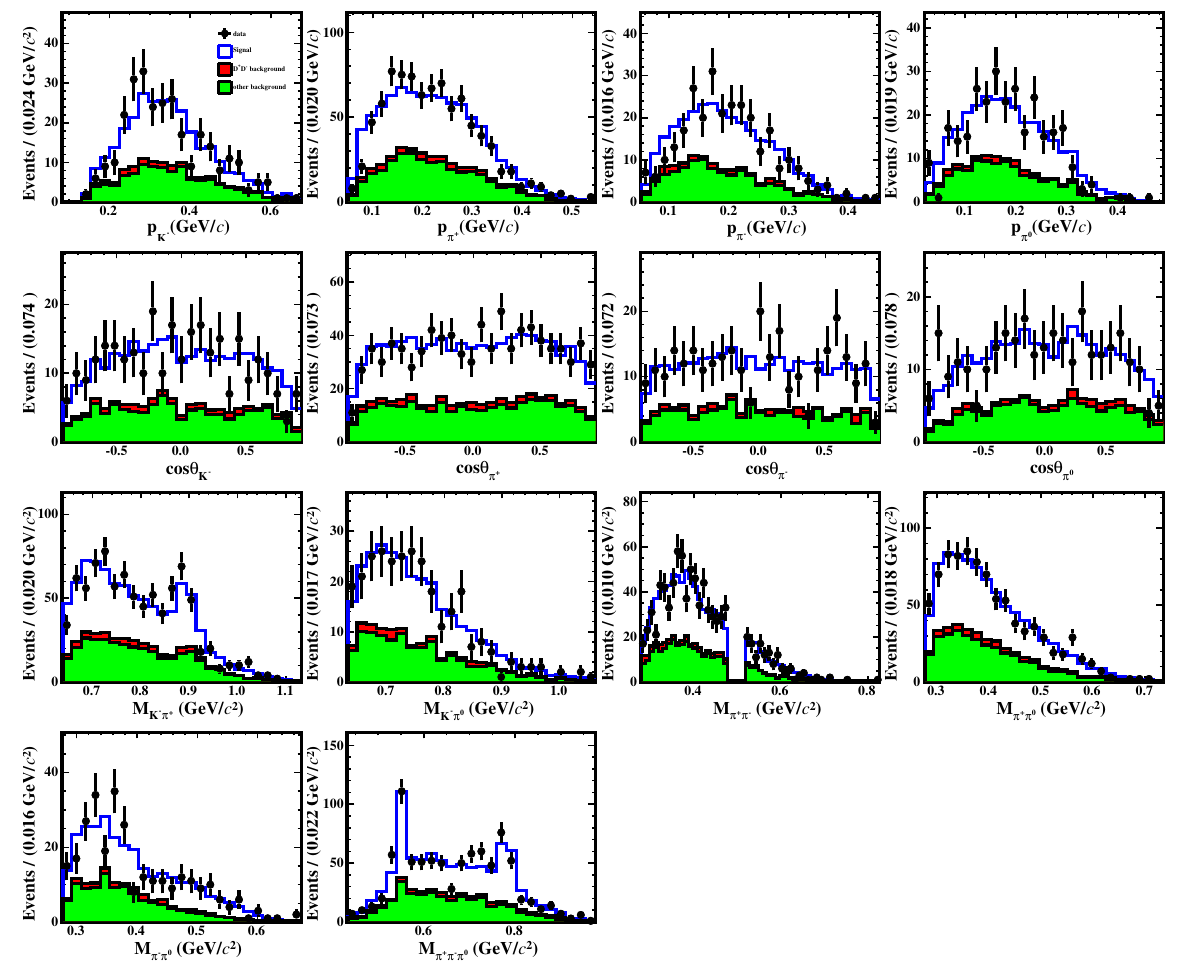}
	\caption{
		Comparisons of the distributions of momenta and cosines of polar angles of daughter particles,
		and the invariant masses of two-body or three-body particle combinations of the accepted candidates for $D^+\to K^- 3\pi^+\pi^-\pi^0$
		between data (dots with error bars) and the total MC simulation (blue histogram). This total MC histogram is the sum of the signal MC events (white histogram) plus the MC-simulated backgrounds from the inclusive MC samples (red and green histograms).
		Events here satisfy the additional requirements of $|M_{\rm BC}^{\rm tag}-1.869|<0.005$~GeV/$c^2$ and $|M_{\rm BC}^{\rm sig}-1.869|<0.005$~GeV/$c^2$
		and there are three entries for $\pi^+$ per event.
		\label{fig:compare3}
	}
\end{figure*}

The DT efficiencies are estimated based on MC simulation. To account for the effect of intermediate resonance structures on the efficiency, each of these decays is modeled by a corresponding mixed-signal MC sample, in which the dominant decay modes containing resonances of
$\eta$, $\omega$, and $K^*(892)$ are mixed with the phase-space signal MC samples. The mixing ratios are determined by
examining the corresponding invariant mass and momentum distributions.
Figures~\ref{fig:compare1}, \ref{fig:compare2}, and \ref{fig:compare3} show
the distributions of momenta and cosines of polar angles of daughter particles,
the invariant masses of two-body or three-body particle combinations of
the accepted candidates for $D^0\to K^- 3\pi^+2\pi^-$, $D^0\to K^- 2\pi^+\pi^-2\pi^0$ and $D^+\to K^- 3\pi^+\pi^-\pi^0$ between data and MC simulations.
Good consistency between data and MC simulation ensures the reliability of the signal efficiencies.

The $\Delta E_{\rm sig}$ requirements,the fitted DT yields~($N_{\rm DT}$), the signal efficiencies ($\epsilon_{\rm sig}$), and the obtained BFs (${\mathcal B}_{\rm sig}$) for each signal decay are summarized in Table~\ref{tab:DT}.

\begin{table*}[htbp]
	\centering
	\caption{\small
		The $\Delta E_{\rm sig}$ requirements,
		the fitted DT yields~($N_{\rm DT}$),
		the signal efficiencies ($\epsilon_{\rm sig}$),
		and the obtained BFs (${\mathcal B}_{\rm sig}$) for each signal decay,
		where the first and second uncertainties of ${\mathcal B}_{\rm sig}$ are statistical and systematic, respectively,
		while the uncertainties of $N_{\rm DT}$ and $\epsilon_{\rm sig}$ are statistical only.
	}\label{tab:DT}
	\begin{tabular}{lcccc}
		\hline\hline
		Signal decay& {\hskip 0.5cm $\Delta E_{\rm sig}$~(GeV) \hskip 0.3cm} & {\hskip 0.3cm $N_{\rm DT}$ \hskip 0.3cm }   & $\epsilon_{\rm sig}$~(\%)    &  ${\mathcal B}_{\rm sig}$~($\times 10^{-4}$) \\ \hline
		$D^0\to K^- 3\pi^+2\pi^-$  &$(-0.031,0.028)$    &$\ \,64\pm11$&$7.20\pm0.19$&$\ \,1.35\pm0.23\pm0.08$\\
		$D^0\to K^- 2\pi^+\pi^-2\pi^0$ &$(-0.042,0.032)$&$441\pm26$&$3.61\pm0.14$&$\ \,19.0\pm1.1\pm1.5$\\
		$D^+\to K^- 3\pi^+\pi^-\pi^0$  &$(-0.036,0.030)$&$157\pm17$&$5.68\pm0.19$&$\ \,6.57\pm0.69\pm0.33$\\
		\hline\hline
	\end{tabular}
\end{table*}

\section{Systematic uncertainties}
\label{sec:sys}

In the measurements of the BFs using Eq.~\ref{eq:br}, all uncertainties associated with the selection of the tag side cancel. Other un-cancelled systematic uncertainties in the BF measurements are discussed below and are stated relative to the measured BFs.

The systematic uncertainties in the total yields of ST $\bar D$ mesons due to the fits to the $M^{\rm tag}_{\rm BC}$ distributions are estimated
to be 0.3\% for the ST $\bar D^0$ and $D^-$~\cite{STyields}.

The systematic uncertainties in tracking or PID of charged tracks are assigned as 0.5\% per $K^\pm$ or $\pi^\pm$,
by using the DT hadronic events with $D$ decaying into $K^-\pi^+$, $K^-\pi^+\pi^+\pi^-$, $K^-\pi^+\pi^+$, and $\bar D$ decaying into their conjugated modes, in which a $K^\pm$ or $\pi^\pm$ is missed, as control samples~\cite{STyields}.

The systematic uncertainty due to $\pi^0$ reconstruction is estimated to be 2.0\% per $\pi^0$, by analyzing
the DT hadronic events of $\bar D^0\to K^+\pi^-\pi^0$ and $\bar D^0\to K^0_S\pi^0$ tagged by either $D^0\to K^-\pi^+$ or $D^0\to K^-\pi^+\pi^+\pi^-$.

The systematic uncertainties of the $\Delta E_{\rm sig}$ requirement are assigned by using control samples of $D^0\to K^-\pi^+\pi^+\pi^-$, $D^0\to K^-\pi^+\pi^0\pi^0$ and $D^+\to K^-\pi^+\pi^+\pi^0$. The differences of the signal efficiencies are taken as the systematic uncertainties, 1.5\%, 3.6\%, and 1.6\% for $D^0\to K^- 3\pi^+2\pi^-$, $D^0\to K^- 2\pi^+\pi^-2\pi^0$, and $D^+\to K^- 3\pi^+\pi^-\pi^0$, respectively.

The systematic uncertainty due to the $K^0_S$ rejection is
studied with the control samples of $D^{+} \to K_{S}^{0} e^+ \nu_{e}$
and $D^{0} \to K_{S}^{0} \pi^+\pi^+\pi^-\pi^-$.
The fitted $K_S^0$ mass and resolutions of data and MC simulation
are in agreement with each other.
This systematic uncertainty is therefore deemed negligible.

The systematic uncertainties due to the mixing MC model are
assigned by varying the fractions of various signal components
by $\pm 1\sigma$ of the quoted BFs, when available; and by one quarter of the component fraction for each of unmeasured processes. The largest changes of the signal efficiencies are assigned as the corresponding
systematic uncertainties, which are
1.8\%, 2.6\%, and 1.2\% for $D^0\to K^- 3\pi^+2\pi^-$, $D^0\to K^- 2\pi^+\pi^-2\pi^0$ and $D^+\to K^- 3\pi^+\pi^-\pi^0$, respectively.

The uncertainties due to MC statistics,
which are 1.1\%, 1.6\%, and 1.3\% for $D^0\to K^- 3\pi^+2\pi^-$, $D^0\to K^- 2\pi^+\pi^-2\pi^0$~ and ~$D^+\to K^- 3\pi^+\pi^-\pi^0$, ~respectively,
are taken as individual systematic uncertainties.

The uncertainty of the quoted BF of $\pi^0\to \gamma\gamma$ is 0.03\% per $\pi^0$~\cite{pdg}.

The systematic uncertainty in the 2D fit to the $M_{\rm BC}^{\rm tag}$ versus $M_{\rm BC}^{\rm sig}$ distribution is examined
in two ways.  An alternative signal shape is obtained by varying the mean and width of the smeared Gaussian resolution function by $\pm 1 \sigma$
and an alternative background shape is obtained by varying the endpoint of the ARGUS function by $\pm0.2$~MeV/$c^2$.
Adding the largest effect from each pair of plus-minus variations
in quadrature gives the systematic uncertainties,
which are 3.5\%, 3.5\%, and 1.5\% for
$D^0\to K^- 3\pi^+2\pi^-$, $D^0\to K^- 2\pi^+\pi^-2\pi^0$ and $D^+\to K^- 3\pi^+\pi^-\pi^0$, respectively.

Table~\ref{tab:relsysuncertainties1} summarizes these systematic uncertainties in the BF measurements.
For each signal decay, the total systematic uncertainty is calculated by adding all above sources in quadrature;
and they are 6.1\%, 7.6\%, and 4.9\% for $D^0\to K^- 3\pi^+2\pi^-$, $D^0\to K^- 2\pi^+\pi^-2\pi^0$ and $D^+\to K^- 3\pi^+\pi^-\pi^0$, respectively.

\begin{table*}[htbp]
	\centering
	\caption{
		Relative systematic uncertainties in \% for the BF measurements.}
	\label{tab:relsysuncertainties1}
	\centering
	\begin{tabular}{lccc}
		\hline\hline
		
		Source           & $D^0\to K^-3\pi^+2\pi^-$& $D^0\to K^-2\pi^+\pi^-2\pi^0$ & $D^+\to K^-3\pi^+\pi^-\pi^0$  \\
		\hline
		$N_{\rm ST}^{\rm tot}$      &0.3    &0.3      &0.3  \\
		$K^-, \pi^\pm$ tracking      &3.0    &2.0      &2.5  \\
		$K^-, \pi^\pm$ PID           &3.0    &2.0      &2.5  \\
		$\pi^0$ reconstruction      &...     &4.0      &2.0  \\
		$\Delta E_{\rm sig}$ cut    &1.5    &3.6      &1.6  \\
		$K_{S}^{0}$ rejection       &Negligible&Negligible&Negligible\\
		MC generator                &1.8    &2.6      &1.2  \\
		MC statistics               &1.1    &1.6      &1.3  \\
		$\pi^0$ $\mathcal B$         &...     &0.06     &0.03 \\
		2D fit                      &3.5    &3.5      &1.5  \\
		\hline
		Total                       &6.1    &7.7      &5.0  \\
		\hline\hline
	\end{tabular}
\end{table*}

\section{Summary}

By analyzing $7.9$~fb$^{-1}$ of $e^+e^-$ collision data taken  with the BESIII detector at $E_{\rm cm}=3.773$~GeV,
we report the first observations of the hadronic decays $D^0\to K^- 2\pi^+\pi^-2\pi^0$ and $D^+\to K^- 3\pi^+\pi^-\pi^0$
as well as an improved measurement of $D^0\to K^- 3\pi^+2\pi^-$.
Their absolute BFs are determined to be
${\mathcal B}(D^0\to K^- 3\pi^+2\pi^-)=( 1.35\pm  0.23\pm  0.08)\times 10^{-4}$,
${\mathcal B}(D^0\to K^- 2\pi^+\pi^-2\pi^0)=( 19.0\pm  1.1\pm  1.5)\times 10^{-4}$, and
${\mathcal B}(D^+\to K^- 3\pi^+\pi^-\pi^0)=( 6.57\pm  0.69\pm  0.33)\times 10^{-4}$,
where the first uncertainties are statistical and the second systematic.
A summary of the measured BFs and the one previous result is shown in Table~\ref{tab:com}.
The results are beneficial for the understanding of the isospin statistical model as applied to the $D^{0(+)}\to K5\pi$ system.
In the near future, amplitude analyses of these decays with a larger data sample from BESIII~\cite{bes3-white-paper}\cite{Li:2021iwf}
will allow exploration of the intermediate states in these decays,
which benefit the understanding of the decay mechanisms of charmed mesons.
\begin{table*}[htbp]
	\centering
	\caption{\small	
		Comparison of the newly-measured BFs and the one previous result; the first and second uncertainties are statistical and systematic, respectively.}
	\label{tab:com}
	\begin{tabular}{lcc}
		\hline\hline
		{Signal decay \hskip 1.8cm} &${\mathcal B}_{\rm sig}(\times 10^{-4})$ & {\hskip 0.5 cm ${\mathcal B}_{\rm FOCUS}(\times 10^{-4})$ \hskip 0.5cm}    \\ \hline
		$D^0\to K^- 3\pi^+2\pi^-$  &$ 1.35\pm  0.23\pm  0.08$&$ 2.2\pm  0.5 \pm 0.3$\\
		$D^0\to K^- 2\pi^+\pi^-2\pi^0$ &$ 19.0\pm  1.1\pm  1.5$&...\\
		$D^+\to K^- 3\pi^+\pi^-\pi^0$  &$ 6.57\pm  0.69\pm  0.33$&...\\
		\hline\hline
	\end{tabular}
\end{table*}
\clearpage
\section{Acknowledgement}

The BESIII Collaboration thanks the staff of BEPCII and the IHEP computing center for their strong support. This work is supported in part by National Key R\&D Program of China under Contracts Nos.  2023YFA1606000; National Natural Science Foundation of China (NSFC) under Contracts Nos. 11635010, 11735014, 11935015, 11935016, 11935018, 12025502, 12035009, 12035013, 12061131003, 12192260, 12192261, 12192262, 12192263, 12192264, 12192265, 12221005, 12225509, 12235017, 12361141819; the Chinese Academy of Sciences (CAS) Large-Scale Scientific Facility Program; the CAS Center for Excellence in Particle Physics (CCEPP); Joint Large-Scale Scientific Facility Funds of the NSFC and CAS under Contract No. U1832207; CAS under Contract No. YSBR-101; 100 Talents Program of CAS; The Institute of Nuclear and Particle Physics (INPAC) and Shanghai Key Laboratory for Particle Physics and Cosmology; Agencia Nacional de InvestigaciÃ³n y Desarrollo de Chile (ANID), Chile under Contract No. ANID PIA/APOYO AFB230003; German Research Foundation DFG under Contract No. FOR5327; Istituto Nazionale di Fisica Nucleare, Italy; Knut and Alice Wallenberg Foundation under Contracts Nos. 2021.0174, 2021.0299; Ministry of Development of Turkey under Contract No. DPT2006K-120470; National Research Foundation of Korea under Contract No. NRF-2022R1A2C1092335; National Science and Technology fund of Mongolia; National Science Research and Innovation Fund (NSRF) via the Program Management Unit for Human Resources \& Institutional Development, Research and Innovation of Thailand under Contract No. B50G670107; Polish National Science Centre under Contract No. 2019/35/O/ST2/02907; Swedish Research Council under Contract No. 2019.04595; The Swedish Foundation for International Cooperation in Research and Higher Education under Contract No. CH2018-7756; U. S. Department of Energy under Contract No. DE-FG02-05ER41374

\end{document}